\documentclass{ws-procs961x669}

\usepackage{dcolumn}
\usepackage{bm}
\usepackage[applemac]{inputenc}
\usepackage[spanish,catalan,english]{babel}
\usepackage{amsmath,amssymb,amsfonts,latexsym,cancel}
\usepackage{graphicx}
\usepackage{color}
\usepackage{soul}
\usepackage{hyperref}
\usepackage{slashed}

\allowdisplaybreaks

\newcommand{\be}{\begin{equation}}
\newcommand{\ee}{\end{equation}}
\newcommand{\bea}{\begin{eqnarray}}
\newcommand{\eea}{\end{eqnarray}}

\begin{document}
\title{{\bf Renormalization and decoupling for the Yukawa model in curved spacetime}}

\author{Sergi Nadal-Gisbert, Antonio Ferreiro  and Jos\'e Navarro-Salas}

\address{Departamento de F\'isica Te\'orica and IFIC, Centro Mixto Universidad de Valencia-CSIC. Facultad de F\'isica, Universidad de Valencia, Burjassot-46100, Valencia, Spain.\\
E-mail: antonio.ferreiro@ific.uv.es, sergi.nadal@uv.es, jnavarro@ific.uv.es}

\begin{abstract}
We consider the renormalization of the one-loop effective action for the Yukawa interaction. We compute the beta functions in the generalized DeWitt-Schwinger subtraction scheme. For the quantized scalar field we obtain that all the beta functions exhibit decoupling for heavy fields as stated by the Appelquist-Carazzone theorem including also the gravitational couplings. For the quantized Dirac field, decoupling appears for almost all of them. We obtain the atypical result that the mass parameter of the background scalar field, does not decouple.\footnote{Expanded version of the talk given by S. Nadal-Gisbert in the Sixteenth Marcel Grossmann Meeting (2021).} \\
\end{abstract}


\bodymatter

\section{Introduction}

Regularization and renormalization of infinities of quantum fields in curved spacetime is a subtle subject. Infinite quantities emerge at the vacuum level and can not be erased due to the non-trivial structure of the classical gravitational background field. There are well established covariant and pragmatic methods to evaluate the vacuum expectation value of the stress-energy tensor in physically reasonable states \cite{birrell-davies, fulling, Waldbook, parker-toms, hu-verdaguer, buchbinder-shapiro}. For a general spacetime, the DeWitt-Schwinger technique \cite{schwinger51,DeWittbook, dewitt75} allows us to evaluate and renormalize the stress-energy tensor as well as the one-loop effective action which constitute a practical tool that encodes all the relevant information about the quantum effects of fields.

 The regularization machinery usually incorporates an arbitrary, non physical, mass-scale parameter. Studying how the coupling constants depend on this scale, i.e., the running of the couplings, is very useful in physical situations where more formal computations are very involved.  This is done by requiring that physical quantities should not depend on spurious parameters \cite{qftbook1,qftbook2,qftbook3}. As a representative case, in dimensional regularization an arbitrary mass scale $\mu$ is introduced to compensate the fictitious dimensions. In the case of the DeWitt-Schwinger proper-time expansion   a mass scale $\mu$ is usually introduced in the short distance logarithmic term $\log \mu^2 \sigma/2$ to overcome an infrared divergence for massless fields. Changing the mass scale $\mu$ to $\mu'$ allows to obtain an effective running for the parameters of the Lagrangian.

Another important feature of effective field theories is that massive fields should decouple at low energies \cite{Babic2002,Sola-Shapiro} as stated by the Appelquist-Carazzone theorem \cite{APtheorem}\footnote{There are some situations were decoupling is violated. This typically happens on theories with spontaneously broken gauge symmetries \cite{Collins,Feruglio}.}. This idea which lies at the heart of the effective field theories states that a field of mass $m$ can not influence the physics at scales larger than $m^{-1}$. In a physical renormalization scheme this idea should be manifested in the beta functions of the theory if they want to describe both infrared and ultraviolet regimes. Moreover, in gravitational physics decoupling is crucial to get a proper physical interpretation in the cosmic infrared regime. This can be particularly relevant for the cosmological constant problem and for the running of the Newton's gravitational constant  \cite{Carroll, Martin, Sola, review}. A partial list of works dealing with this issue in a curved space are \cite{franchinoreview,franchino19,gorbar-shapiro2003,gorbar-shapiro2004, antipin, markkanen}. 

Recently, an arbitrary scale parameter $\mu$ has been introduced in the DeWitt-Schwinger subtraction scheme in order to consistently construct the subtraction terms avoiding the infrared divergences \cite{FN20} (see also \cite{FN, moreno-sola, BDNN}). This was done for the free complex scalar field in the Einstein-Maxwell theory. This approach has the benefit of producing the decoupling of heavy fields. In this work we further extend this approach by including Yukawa interactions with spinor fields and scalars also coupled to gravity. This model has been broadly studied in dimensional regularization \cite{Toms18,barra-buchbinder19,buchbinder19,gorbar-shapiro2004} also by including gauge fields \cite{toms18gauge, toms19}. We will concentrate in the simplest case of a scalar background coupled to a quantized dirac field and also to a scalar field. The aim of this work is to compute the beta functions in the generalized DeWitt-Schwinger renormalization approach and observe whether decoupling explicitly appears.

  \section{Interaction with a quantized scalar field and renormalization}\label{SectionScalars}
Consider a quantized real scalar field $\varphi$ coupled to a real scalar background $ \phi$ via the Yukawa interaction $\frac{h^2}{2}\phi^2\varphi^2$

\begin{align}\label{ClassicalAction}
S = \int d^4x \sqrt{-g} & \left\{ -\Lambda + \frac{R}{16\pi G} + \frac12 \nabla^{\mu}\varphi\nabla_{\mu}\varphi - \frac12 \left(m^2 +\xi R \right) \varphi^2 \right. \nonumber \\
& \left. - \frac{h^2}{2}\phi^2\varphi^2  + \frac12 \nabla^{\mu}\phi\nabla_{\mu}\phi - V(\phi) \right\} \ ,
\end{align}
where $m^2$ is the mass parameter for the quantized scalar, $\xi$ is the coupling of $\varphi^2$ to the Ricci scalar, and $V(\phi)$ is a general potential that can contain interactions between the background field and the curvature but is independent of the quantized scalar field $\varphi$. The Feynman propagator $G_F$ for $\varphi$ satisfies
\be\label{ClassicalLagrangian1} \left(\Box_x+m^2+\xi R+h^2\phi^2\right)G_{\rm F}(x,x')=-|g(x)|^{-1/2}\delta(x-x').\ee
The effective action can be generated from this propagator by $S_{\rm eff}=-i \frac12 \operatorname{Tr} \log{(-G_{\rm F})}$. The ultraviolet divergences of the one-loop effective Lagrangian can be explicitly manifested if one expresses the Feynman propagator as an integral in the proper time $s$,
\be \label{GFs} G_{\rm F}(x, x') = -i \int_0^\infty ds \ e^{ -im^2 s} \langle x, s |  x', 0\rangle \ , \ee 
where $m^2$ is understood to have an infinitesimal negative imaginary part ($m^2\equiv m^2 -i\epsilon$). 
The heat kernel $\langle x, s |  x', 0\rangle$
can be expanded in powers of the proper time as follows 
 {\be \label{hks}\langle x, s |  x', 0\rangle =  i\frac{\Delta^{1/2} (x, x')}{(4\pi)^2(is)^2}  \exp {\frac{\sigma(x, x')}{2is}} \sum_{j=0}^\infty a_j (x, x')(is)^j \ , \ee 
where $\Delta(x, x')$ is the Van Vleck-Morette determinant and $\sigma(x, x')$ is  the proper distance along  the geodesic from $x'$ to $x$. Therefore, the effective Lagrangian, defined as $S_{\rm eff}=\int d^4x \sqrt{-g}L_{\rm eff}$, has the following  asymptotic expansion 
\be
L_{\rm eff}=\frac{i}{2(4\pi)^{2}}\sum^{\infty}_{j=0}a_j(x)\int^{\infty}_0e^{-is m^2}(is)^{j-3}ds \label{Leff}\ .
\ee
The first coefficients $a_n(x, x')$ are given, in the coincidence limit $x \to x'$, by  \cite{parker-toms}
\begin{align}
a_0(x)=&1 \ ,  \ \ \ \  \ \ a_1(x)=\frac16R-Q \nonumber\\
a_2(x)=&\frac{1}{180}R_{\alpha\beta\gamma\delta}R^{\alpha\beta\gamma\delta}-\frac{1}{180}R^{\alpha\beta}R_{\alpha\beta} -\frac{1}{30}\Box R+\frac{1}{72}\ R^2 \nonumber \\
&+\frac12 Q^2-\frac16 R Q +\frac16 \Box Q + \frac{1}{12} W_{\mu\nu}W^{\mu\nu}\ , \label{coef} 
 \end{align}
where for the case of a scalar field
we have $W_{\mu\nu}\equiv \left[\nabla_{\mu}, \nabla_{\nu} \right] = 0$ and $Q=\xi R+h^2 \phi^2$. Expansion \eqref{Leff} shows that the ultraviolet divergent terms of the effective Lagrangian are localized in the first three terms of the DeWitt-Schwinger expansion in the limit $s\to0$ of the integral. The renormalization procedure can be performed by directly subtracting these divergent terms to the total one-loop contribution. 
\be
L_{\rm ren} =\int^{\infty}_0 ds\left[\frac{e^{-im^2s}}{is}\langle x,s|x,0 \rangle  - \frac{i}{2(4\pi)^{2}}\sum^{2}_{j=0}a_j(x)\frac{e^{-is m^2}}{(is)^{-j+3}}\right] \ . \label{ren}
\ee
where the notation of $L_{\rm ren}$ has to be understood as the finite one-loop correction to the background Lagrangian. One can notice that the massless case inherits an infrared divergence ($s \to \infty$ limit). We avoid it by introducing a mass scale $\mu^2$ in the exponential term of the DeWitt-Schwinger expansion by writing $\sum_j a_j(x) e^{-ism^2} \to \sum_j \bar a_j(x) e^{-is(m^2+ \mu^2)}$ in (\ref{Leff}). As mentioned in \cite{FN20}, this is the unique way of introducing the $\mu$ parameter if we want the exponential form to remain. The main point of the introduction of the arbitrary parameter $\mu$ at the same level of the mass, is to obtain decoupling in the infrared behaviour of the beta functions, as we will see. The DeWitt coefficients $a_j$ are redefined by consistency with the adiabatic orders, $\bar{a}_0(x)=1, \bar{a}_1(x)=a_1(x)+\mu^2, \bar{a}_2(x)=a_2(x)+ a_1(x)\mu^2+\frac12\mu^4$, and renormalization is now performed by taking the difference $L_{\rm ren}(\mu)= L_{\rm eff} -L_{\rm div}(\mu)$, as for the standard situation with $\mu=0$, \eqref{ren}

\be
L_{\rm ren}(\mu)=\int^{\infty}_0 ds\left[\frac{e^{-im^2s}}{is}\langle x,s|x,0 \rangle  - \frac{i}{2(4\pi)^{2}}\sum^{2}_{j=0}\bar{a}_j(x)\frac{e^{-is(m^2 + \mu^2)}}{(is)^{-j+3}}\right]  \label{renmu}
\ee
We note that the renormalized one-loop Lagrangian depends now on $\mu$. This dependece in the arbitrary parameter $\mu$ has to be compensated by a running of the couplings of the background Lagrangian, thus the physical one-loop renormalized Lagrangian $L_{\rm phys} = L_{\rm B}(\mu)+ L_{\rm ren} (\mu)$ is $\mu$-independent, where $L_{\rm B}(\mu)$ refers to the background Lagrangian. 

\subsection{Running of the coupling constants and decoupling}
The background Lagrangian required for renormalization is given by
\be\label{ClassicalLagrangian1}
L_{\rm B}= L_{\rm grav} +\frac{1}{2}Z \nabla^{\mu}\phi\nabla_{\mu}\phi - \frac{M^2}{2} Z \phi^2 -\frac{\xi_{\phi}}{2} R Z\phi^2  - \frac{\lambda}{4!}Z^2\phi^4  + \gamma_1 \Box Z\phi^2 \ ,
\ee
where
\be\label{GravitationalLagrangian}
L_{\rm grav} = -\Lambda +\frac{1}{2} \kappa R + \alpha_1R^2 
+ \alpha_2 R_{\mu\nu}R^{\mu\nu} + \alpha_3 R_{\mu\nu\alpha\beta}R^{\mu\nu\alpha\beta}+ \alpha_4  \Box R \ .
\ee
We have defined $\kappa= 1/8\pi G$. The new terms are required to absorb the divergences of the one-loop correction. The coupling $Z$ will not receive any contribution from the scalar quantum fluctuations, therefore we can canonically normalize it to 1. The remaining couplings $\lambda(\mu), \kappa(\mu), \alpha_i(\mu), M(\mu)$, etc will inherit a dependence  on the mass scale $\mu$.

The beta functions and the running couplings are built from the renormalization process, therefore they get only contributions from the divergent terms. For this reason, it is enough to approximate \eqref{renmu} with the first three terms of the asymptotic expansion for the heat kernel.

\begin{align}
L_{\rm ren}(\mu) \eqsim \frac{i}{2(4\pi)^{2}}\int^{\infty}_0 \frac{ds}{s^3} & \left\{e^{-im^2s}\left[1 + (i s)a_1(x) + (is)^2 a_2(x) \right] \right. \nonumber \\
& \left. - e^{-is(m^2 + \mu^2)}\left[1 + (i s)\bar{a}_1(x) + (is)^2 \bar{a}_2(x) \right]\right\} \ . \label{renmu2}
\end{align}
This is a finite integral that can now be computed. Its result depends on the DeWitt coeficients $a_0(x)$, $a_1(x)$ and $a_2(x)$ and also with the arbitrary $\mu$ parameter. Requiring that the physical one-loop renormalized Lagrangian $L_{\rm phys} = L_{\rm B}(\mu)+ L_{\rm ren}(\mu) $ has to be $\mu$-independent, leads to the running of the couplings and the beta functions. 

\be\label{RunningsEq}
\frac{d L_{\rm phys}}{d \mu} = 0 \quad \rightarrow \quad \beta_i \frac{\partial L_{B}}{\partial q_i} = -\mu \frac{\partial L_{\rm ren}}{\partial \mu} \ .
\ee
where $q_i$ refers to all the parameters of the Lagrangian and $\beta_i = \mu \frac{\partial q_i}{\partial \mu}$. A representative of the beta functions that we obtain are 
\begin{alignat}{2}\label{betaDS1ScalarsDimenLess}
\beta_{\lambda} & =\frac{3 h^4}{4\pi^2}\frac{\mu^2}{m^2+\mu^2} &\qquad
\beta_{\Lambda} & =\frac{1}{32\pi^2}\frac{\mu^6}{m^2+\mu^2} \nonumber \\
\beta_{\kappa} & =\frac{\bar\xi}{8\pi^2}\frac{\mu^4}{m^2+\mu^2}& \qquad
\beta_{M^2}& =-\frac{h^2}{8\pi^2}\frac{\mu^4}{m^2+\mu^2}  \ ,
\end{alignat}
where we have defined $\bar\xi=\left(\xi-\frac16\right)$. The rest of beta functions can be found in Appendix A.  Several remarks are worth of mention in these results. For the dimensionless parameters, like $\lambda(\mu)$, we obtain a factor of the form $\mu^2/(m^2 + \mu^2)$, this factor also arises in the hierarchy of beta functions in the Wilsonian renormalization approach for a scalar field theory \cite{Hollowood}. Analyzing this factor it is easy to see that for $m^2 \gg \mu^2$ the dimensionless beta functions decouple as expected in the infrared regime. Moreover for $\mu^2 \gg m^2$ the ultraviolet regime is recovered as expected from dimensional regularization with MS (minimal subtraction) \cite{Toms18}. Another important feature concerns the dimensional couplings. For the Newton and cosmological constants we  recover the same result as in the free field theory. Furthermore, all the dimensional couplings also decouple when $m^2\gg \mu^2$.

\section{Renormalization for Dirac fields}\label{SectionDirac}

In this section we consider a quantized Dirac field $ \psi$ coupled via Yukawa interaction $g_Y\phi\bar\psi\psi$ with a classical scalar background $\phi$. The action is given by
\be\label{YukawaFermionsClassicalAction}
S=\int d^4x \sqrt{-g}\left( -\Lambda + \frac{R}{16\pi G} + \bar\psi \left( i\gamma^{\mu}\nabla_{\mu} - m\right)\psi - g_Y\phi\bar\psi\psi   + \frac12 \nabla^{\mu}\phi\nabla_{\mu}\phi - V(\phi) \right) 
\ , \ee
where the covariant derivative $\nabla_\mu$ acting on the Dirac field  is defined as the  ordinary derivative plus the spin connection term. $\gamma^\mu(x)$ are the curved space Dirac matrices $\gamma^\mu(x)= e^{\ \mu}_a \gamma^a$, defined in terms of the usual Dirac matrices in Minkowski space $\gamma^a$ and the vierbein $e^{\ \mu}_a$. 

In order to build the DeWitt-Schwinger expansion one needs the Klein-Gordon equation for the Dirac field
   \be \label{KG1Ferm}(\Box_x  + m^2 + Q) G_{\rm F}(x, x') = -|g(x)|^{-1/2} \delta (x-x') \ . \ee
 Consequently, the one-loop effective action takes the form $S_{\rm eff}^{(1)}= \frac{1}{2}i \operatorname{Tr} \log (-G_F) $.  One can expand the heat kernel in the proper time asymptotic series as for the scalar field \eqref{hks} and make use of the same coefficients \eqref{coef}. In this case $Q$ is given by
\be Q(x) = \frac14 R(x) + i g_Y \gamma^{\mu}\nabla_{\mu}\phi(x) + g_Y^2 \phi^2 (x)+ 2 g_Y m \phi(x)  \label{qfermion}\ . \ee
%
%
It is important to stress the appearance of the last term in the above expression which is proportional to $m$. This contrast with the obtained expression for $Q$ in the scalar case ($Q=\xi R+h^2 \phi^2$). Generically, all DeWitt coefficients  $a_n$ (or $\bar a_n$) are local geometrical quantities independent of the mass of the field. The Yukawa interaction for Dirac fermions introduces a mass-dependent term in the expression for $Q(x)$. This term will be the responsible of the violation of the decoupling characteristic for the background scalar mass parameter.

As in the previous section one gets the expression for the subtraction terms with the asymptotic expansion
\be
L_{\rm div}(\mu)=\frac{-i}{2(4\pi)^{2}}\sum^{2}_{j=0}\operatorname{tr}\bar{a}_j(x)\int^{\infty}_0e^{-is(m^2 + \mu^2)}(is)^{j-3}ds \ , \label{Ldifmu3}
\ee
The main difference with the scalar case is that one needs to take into account the trace of the spinor indices acting on the coefficients $\operatorname{tr}\bar{a}_j(x)$. We still have the modified DeWitt coefficients $\bar{a}_0(x)=1$, $\bar{a}_1(x)=a_1(x)+\mu^2$,  and $\bar{a}_2(x)=a_2(x)+\left(\frac{1}{6}R - Q\right)\mu^2+\frac12\mu^4$
related to the DeWitt coefficients  \eqref{coef} but now with $Q$ given by \eqref{qfermion}  and $W_{\mu\nu} = [\nabla_{\mu},\nabla_{\nu}]= -\frac{1}{8} R_{\mu\nu a b}\left[\gamma^a, \gamma^b \right] $.\\

\subsection{Running of the coupling constant and decoupling}
The background Lagrangian terms are required by renormalization in order to absorb the divergent terms coming from the one-loop correction \eqref{Ldifmu3}
\begin{align}\label{ClassicalL}
L_{\rm B} & = L_{\rm grav} +\frac{1}{2} Z \nabla^{\mu}\phi\nabla_{\mu}\phi - \frac{M^2}{2} Z \phi^2 
-\frac{\xi}{2} R Z\phi^2 -\tau Z^{1/2}\phi - \frac{\eta}{3!}Z^{3/2}\phi^3 \nonumber\\
&- \frac{\lambda}{4!}Z^2\phi^4 - \xi_1R Z^{1/2}\phi + \gamma_1 \Box Z\phi^2 +  \gamma_2\Box Z^{1/2} \phi \ ,
\end{align}
where $L_{\rm grav}$ was given in \eqref{GravitationalLagrangian}. In the Dirac case, $Z$ gets a contribution from the quantum fluctuations of the Dirac field. Therefore, it has a running that can be related to a running of the field $\phi$ by a reparametrization. We write $Z = 1 + \delta Z $ as usual, to take into account canonical normalization and the one-loop correction. For simplicity, we introduce new primed couplings to absorb $Z$, except for the kinetic term, where we leave $Z$ explicitly
\be\label{ClassicalLPrimed}
 \frac{1}{2} Z\nabla^{\mu}\phi\nabla_{\mu}\phi - \frac{M'^2}{2} \phi^2 - \frac{\xi'}{2} R \phi^2 -\tau'\phi - \frac{\eta'}{3!}\phi^3 - \frac{\lambda'}{4!}\phi^4 - \xi'_1R \phi + \gamma'_1 \Box \phi^2 + \gamma'_2\Box  \phi \ .
\ee
As for the scalars, we just need to consider the divergent part of the one-loop effective action \eqref{renmu2} to compute the running of the parameters. Again, we impose that the physical one-loop renormalized Lagrangian $L_{\rm phys}=L_{\rm B}(\mu)+ L_{\rm ren}(\mu)$ must be independent of the value of $\mu$. Therefore considering \eqref{RunningsEq} for this background Lagrangian we can obtain the beta functions for the primed parameters. By direct differentiation and keeping one-loop order $\mathcal{O}(\hbar)$, it is straightforward to obtain the beta functions of the unprimed original couplings 

\begin{alignat}{2}\label{BetaFunctStructureOne-loop}
\beta_{M^2} & = \beta_{M'^2}- M^2\beta_{Z}    & \qquad
\beta_{\xi} & = \beta_{\xi'}- \xi \beta_{Z}  \nonumber \\
\beta_{\xi_1} & = \beta_{\xi'_1}- \frac 12 \xi_1 \beta_{Z} &\qquad
\beta_{\tau} & = \beta_{\tau'}- \frac{\tau}{2}\beta_{Z}  \nonumber \\
\beta_{\eta} &= \beta_{\eta'}- \frac{3}{2}\eta \beta_{Z}  &\qquad
\beta_{\lambda} & = \beta_{\lambda'}- 2\lambda \beta_{Z} \nonumber \\
\beta_{\gamma_1} & = \beta_{\gamma'_1}- \gamma_1\beta_{Z}  &\qquad
\beta_{\gamma_2} & = \beta_{\gamma'_2}- \frac12 \gamma_2\beta_{Z} \ .\nonumber 
\end{alignat}
%
%
The result for all the beta functions  can be seen in Appendix A. Let us analyze the different regimes of the scalar wave function $Z$ as a representative of the dimensionless parameters.
\be
\beta_{Z}=-\frac{g_Y^2}{4\pi^2}\frac{\mu^2}{m^2+\mu^2} \label{betaq2}\ .
\ee
In the ultraviolet regime, $\mu \gg m$ we recover the result from dimensional regularization with MS \cite{Toms18, barra-buchbinder19}.
\be \beta_{Z}=-\frac{g_Y^2}{4\pi^2}\frac{\mu^2}{m^2+\mu^2} \to_{\mu \gg m}  -\frac{g_Y^2}{4\pi^2} \ ,\ee
while for the infrared regime, $\mu \ll m$, we find  decoupling
\be \beta_{Z}=-\frac{g_Y^2}{4\pi^2}\frac{\mu^2}{m^2+\mu^2} \to_{\mu \ll m}  -\frac{g_Y^2}{4\pi^2}\frac{\mu^2}{m^2} \ .\ee
The other beta functions for the dimensionless parameters exhibit the same behaviour. Decoupling of massive fields is manifested in the infrared regime and the ultraviolet regime agrees with the results of dimensional regularization with MS \cite{Toms18,barra-buchbinder19}. 

Things are more involved for the dimensional constants. We observe different behaviours with the parameter $\mu$. For example for the couplings $\Lambda(\mu)$, $\kappa(\mu)$ and $M'^2(\mu)$ we get
\begin{alignat}{2}\label{BetaFermions}
\beta_{\Lambda}&=-\frac{1}{8\pi^2}\frac{\mu^6}{m^2+\mu^2} & \quad
\beta_{\kappa}&=-\frac{1}{24\pi^2}\frac{\mu^4}{m^2+\mu^2} \nonumber  \\
\beta_{M^2}&=\frac{g_Y^2}{8\pi^2}\frac{\mu^2}{m^2 + \mu^2}\left(+4\mu^2-8m^2 + 2M^2 \right)  \ .
\end{alignat}
A common feature is that decoupling appears when $m^2 \gg \mu^2$ for all the dimensional couplings except for the background mass parameter $M^2$. In this case the beta function essentially reproduces the value obtained via dimensional regularization with MS times a factor of order $\mu^2/m^2$. The origin of the term proportional to $-8m^2$ can be retrieved from the term $2g_Y m \phi$ in $Q$.  This can be linked to the fact that when building the DeWitt-Schwinger expansion all the mass dependence is assumed to be encoded in the exponential term \cite{Christensen76} in the form $m^2$ and this expansion does not work properly if $Q$ inherits a mass dependence. The latter result shows that finding decoupling for all the coupling constants, of a given theory, although it is crucial to define a cosmic infrared regime, it is indeed a nontrivial task, as already emphasized in \cite{franchinoreview}.

Finally it is worth showing the finite expression for the running  of the Newton gravitational constant $G$, the cosmological constant $\Lambda$ and the scalar mass term $M'^2$ coming from the Dirac field at two different scales $\mu$ and $\mu_0$.
\bea \label{rlambda}
\Lambda(\mu)&=&\Lambda_0-\frac{1}{32\pi^2}\left((\mu^4-\mu_0^4)-2m^2(\mu^2-\mu_0^2) 
+ 2m^4\log{\left(\frac{m^2+\mu^2}{m^2+\mu_0^2}\right)}\right)\ , \label{ccmu}  \\
G(\mu)&=&\frac{G_0}{1-\frac{ G_0}{6\pi}\left(\mu^2-\mu_0^2-m^2\log{\left(\frac{m^2+ \mu^2}{m^2+\mu_0^2}\right)}\right)} \ , \label{rG} \\
M'^{2}(\mu)&=&M'^{2}_0+\frac{g^2_Y}{4\pi^2}\left((\mu^2-\mu_0^2)-3m^2\log{\left(\frac{m^2+ \mu^2}{m^2+\mu_0^2}\right)}\right)\ ,\label{rM}
\eea
where $G_0$, $\Lambda_0$ and $M'^2_0$ are the parameters at the scale $\mu_0$. Note the appearance of $\mu^4$ and $\mu^2$ terms. In more conventional approaches only the logarithmic terms are present. However these non-logarithmic terms can be traced to the non trivial quartic and quadratic divergences and are indeed responsible of the appearance of decoupling. Expanding $\Lambda(\mu)$ and $G(\mu)$  for $\mu^2 \gg m^2$ one can check that the runnings are suppressed by a factor $\mu^6/m^2$ and $\mu^4/m^2$. However for $M'^2(\mu)$ we still get a non suppressed term proportional to $\mu^2$.\\

\section{Conclusions and final comments}  \label{conclusions} 
We have considered the one-loop effective action for a quantized scalar field and Dirac field coupled to a background scalar field and gravity. Using the upgraded DeWitt-Schwinger subtraction scheme we have renormalized the theory and computed the beta functions for all the parameters. As a remarkable result, for the dimensionless couplings we obtain the beta functions that satisfy both ultraviolet and infrared regimes. In the ultraviolet regime we recover the beta function from MS scheme. In the infrared regime the beta functions explicitly exhibit the decoupling property. For the dimensional parameters new $\mu^4$ and $\mu^2$ terms appear in the runnings. These terms signal the presence of quadratic and quartic divergences and are indeed the responsible for the decoupling in the running couplings. Therefore, the obtained beta functions of the dimensional parameters exhibit the property of decoupling in the low energy regime, including the Newton's and cosmological constant. The exception is for the contribution of the Dirac field to the scalar mass parameter. As stress before the reason is localized in the linear term $2 g_y m \phi$ in Q. \\

{\it Acknowledgments.--} 
This work has been  partially supported  by Spanish Ministerio de  Economia,  Industria  y  Competitividad  Grants  No. FIS2017-84440-C2-1-P (MINECO/FEDER, EU) and No.  FIS2017-91161-EXP, and also by the project PROMETEO/2020/079 (Generalitat Valenciana).  A. F. is supported by the Severo Ochoa Ph.D. fellowship, Grant No. SEV-2014-0398-16-1, and the European Social Fund. S. N. is supported by the Universidad de Valencia, within the Atracci\'o de Talent Ph.D fellowship No. UV-INV- 506 PREDOC19F1-1005367.

\appendix

The obtained beta functions for the scalar one-loop correction are

\begin{alignat}{2}\label{betaDS1ScalarsDimenLess}
\beta_{\xi_{\phi}}&=\frac{h^2\bar\xi}{8\pi^2}\frac{\mu^2}{m^2+\mu^2} &\qquad \beta_{\alpha_1} & =-\frac{\bar\xi^2}{32\pi^2}\frac{\mu^2}{m^2+\mu^2} \nonumber \\
\beta_{\alpha_4} &=-\frac{\xi-\frac15}{96\pi^2}\frac{\mu^2}{m^2+\mu^2} &\qquad
\beta_{\alpha_2} & =\frac{1}{2880\pi^2}\frac{\mu^2}{m^2+\mu^2}\nonumber\\
\beta_{\alpha_3} & =-\frac{1}{2880\pi^2}\frac{\mu^2}{m^2+\mu^2} &\qquad
\beta{\gamma_1} & = -\frac{h^2}{96\pi^2}\frac{\mu^2}{m^2+\mu^2} \nonumber \\
\beta_{\lambda} & =\frac{3 h^4}{4\pi^2}\frac{\mu^2}{m^2+\mu^2} &\qquad
\beta_{\Lambda} & =\frac{1}{32\pi^2}\frac{\mu^6}{m^2+\mu^2} \nonumber \\
\beta_{\kappa} & =\frac{\bar\xi}{8\pi^2}\frac{\mu^4}{m^2+\mu^2}& \qquad
\beta_{M^2}& =-\frac{h^2}{8\pi^2}\frac{\mu^4}{m^2+\mu^2}  \ ,
\end{alignat}
Dimensionless beta functions for the Dirac field are:
\begin{alignat}{2}\label{betaFermionsAdimen}
\beta_{Z}&=-\frac{g_Y^2}{4\pi^2}\frac{\mu^2}{m^2+\mu^2} \nonumber \\
 \beta_{\xi} & =-\frac{g_Y^2}{24\pi^2}\frac{\mu^2}{m^2+\mu^2}\left(1-6\xi\right) &\qquad \beta_{\lambda}&=-\frac{ g_Y^2 }{8\pi^2}\frac{\mu^2}{m^2+\mu^2}\left(24g_Y^2 - 4\lambda\right)    \nonumber\\
 \beta_{\alpha_1}& =\frac{1}{1152\pi^2}\frac{\mu^2}{m^2+\mu^2} &\qquad
\beta_{\alpha_2}&=-\frac{1}{720\pi^2}\frac{\mu^2}{m^2+\mu^2}  \nonumber \\
\beta_{\alpha_3}&=-\frac{7}{5760\pi^2}\frac{\mu^2}{m^2+\mu^2} &\qquad
\beta_{\alpha_4}&= \frac{1}{480\pi^2}\frac{\mu^2}{m^2+\mu^2}  \nonumber \\
\beta_{\gamma_1}& =\frac{g_Y^2}{4\pi^2}\frac{\mu^2}{m^2+\mu^2}\left(\frac16 +\gamma_1\right) &\qquad \beta_{\gamma_2}&=\frac{g_Y}{8\pi^2}\frac{\mu^2}{m^2+\mu^2} \left(\frac23 m + \gamma_2\right) \ .
\end{alignat}
Dimensional beta functions for the Dirac field: 
\begin{alignat}{2}\label{betaFermionsDimensional}
\beta_{\Lambda}&=-\frac{1}{8\pi^2}\frac{\mu^6}{m^2+\mu^2} &\qquad \beta_{\xi_1}&=\frac{g_Y}{8 \pi^2}\frac{\mu^2}{m^2+\mu^2}\left(-\frac{m}{3} + g_Y\xi_1\right) \nonumber \\
\beta_{\kappa}&=-\frac{1}{24\pi^2}\frac{\mu^4}{m^2+\mu^2} &\qquad \beta_{\tau}&=-\frac{g_Y  }{8\pi^2}\frac{\mu^2}{m^2+\mu^2}\left(4m\mu^2 - \tau g_Y \right) \nonumber \\
\beta_{M^2}&=\frac{g_Y^2}{8\pi^2}\frac{\mu^2}{m^2 + \mu^2}\left(+4\mu^2-8m^2 + 2M^2 \right) &\qquad \beta_{\eta}&=-\frac{ g_Y^3 }{8\pi^2}\frac{\mu^2}{m^2+\mu^2} \left(24 g_Y m - 3\eta \right) \ .
\end{alignat}

\end{document}